\documentclass[usenatbib, epsfig, a4paper]{mn2e} 
\pdfoutput=1 

\usepackage{times}
\usepackage{subfigure}
\usepackage{amsmath}
\usepackage{amsfonts}
\usepackage{amssymb}
\usepackage[pdftex]{graphicx} 
\usepackage{multirow}
\usepackage{hyperref}
\usepackage{url}
\usepackage{bm}
\usepackage{ulem}
\usepackage[abs]{overpic}

\renewcommand{\d}[1]{\ensuremath{\operatorname{d}\!{#1}}}
\newcommand {\apgt} {\ {\raise-.5ex\hbox{$\buildrel>\over\sim$}}\ }
\newcommand {\aplt} {\ {\raise-.5ex\hbox{$\buildrel<\over\sim$}}\ }

\newenvironment{packed_enum}{
\begin{enumerate}
  \setlength{\itemsep}{1pt}
  \setlength{\parskip}{0pt}
  \setlength{\parsep}{0pt}
}{\end{enumerate}}

\newcommand\lsim{\mathrel{\rlap{\lower4pt\hbox{\hskip1pt$\sim$}}
        \raise1pt\hbox{$<$}}}
\newcommand\gsim{\mathrel{\rlap{\lower4pt\hbox{\hskip1pt$\sim$}}
        \raise1pt\hbox{$>$}}}
\def\myputfigure#1#2#3#4#5%
{\vskip#5pt\makebox[0pt]{\hskip#2in
\includegraphics[width=#3\textwidth]{#1}}\vskip#4pt\hfill}

\newcommand{\cmfast}{\textsc{\small 21CMFAST}}
\newcommand{\cmfastvone}{\textsc{\small 21CMFASTv1}}

\newcommand{\hi}{H\thinspace\textsc{i} }
\newcommand{\hii}{H\thinspace\textsc{ii} }

\title[21-cm signatures of residual HI]{21-cm signatures of residual HI inside cosmic HII regions during reionization} 
\author[C. A. Watkinson, A. Mesinger, J. R. Pritchard \& E. Sobacchi]
{C.~A.~Watkinson$^1$\thanks{Email: \href{mailto:c.watkinson11@imperial.ac.uk}{\protect\nolinkurl{c.watkinson11@imperial.ac.uk}}}, A. Mesinger$^2$, J.~R.~Pritchard$^1$ \& E. Sobacchi$^2$\\ 
$^1$Department of Physics, Blackett Laboratory, Imperial College, London SW7 2AZ, UK\\
$^2$Scuola Normale Superiore, Piazza dei Cavalieri 7, I-56126 Pisa, Italy
}
 
\date{\today}

\pubyear{2015}

\begin{document}

\def\*#1*\ {} 

\maketitle

\begin{abstract}
We investigate the impact of sinks of ionizing radiation on 
the reionization-era 21-cm signal, focusing on 1-point statistics. 
We consider sinks in both the intergalactic medium and inside galaxies.
At a fixed filling factor of \hii regions, sinks will have two 
main effects on the 21-cm morphology: 
(i) as inhomogeneous 
absorbers of ionizing photons they result in smaller and 
more widespread cosmic \hii patches; 
and (ii) as reservoirs 
of neutral gas they contribute a non-zero 21-cm 
signal in otherwise ionized regions. 
Both effects damp the contrast between neutral 
and ionized patches during reionization, 
making detection of the epoch of reionization 
with 21-cm interferometry more challenging. 
Here we systematically investigate these effects 
using the latest semi-numerical simulations.  
We find that sinks dramatically suppress the 
peak in the redshift evolution of the variance, 
corresponding to the midpoint of reionization. 
As previously predicted, skewness changes sign 
at midpoint, but the fluctuations in the residual 
\hi suppress a late-time rise. 
Furthermore, large levels of residual 
\hi dramatically alter the evolution
of the variance, skewness and power spectrum
from that seen at lower levels.
In general, 
the evolution of the large-scale modes provides a 
better, cleaner, higher signal-to-noise probe 
of reionization.
\end{abstract}

\section{Introduction}

Reionization, the process in which the first galaxies 
ionized their surroundings and ultimately the entire Universe,
is poorly constrained at present. 
The 21-cm signal, as produced by a hyperfine transition
in neutral hydrogen (H\thinspace\textsc{i}),
provides an excellent tracer of the neutral IGM.
It is therefore the hope that
future observations of the 21-cm brightness temperature
using immense radio arrays, e.g. 
the LOw Frequency ARray\footnote{\url{http://www.lofar.org/}} (LOFAR),
the Murchison Wide-field Array\footnote{\url{http://www.mwatelescope.org/}}
 (MWA), 
the Precision Array for Probing the Epoch of Reionization\footnote{\url{http://eor.berkeley.edu/}} (PAPER),
the Hydrogen Epoch of Reionization Array\footnote{\url{http://reionization.org/}} 
(HERA) and 
the Square Kilometre Array\footnote{\url{http://www.skatelescope.org/}} 
(SKA), 
will dramatically improve our understanding of  reionization.

To make optimal use of these observations it
is vital that we build insight into
the wide range of physical processes involved in reionization. 
Two of the most important factors to consider are
the star formation history and
the degree to which ionizing radiation 
from galaxies escapes into the intergalactic medium (IGM). 
These two physical mechanisms
have naturally been the focus in the 
modelling of reionization, with both semi-numerical
(e.g. \citealt{Zahn2007, Mesinger2007, Choudhury2009, Signal2010, Santos2010, Thomas2011})
and numerical simulations 
(e.g. \citealt{Trac2007, Zahn2007, Baek2009, Trac2011, Iliev2013}) 
being used to refine 
our understanding of both.
These simulations describe 
how ionized regions form, 
grow and ultimately merge 
to ionize the entire IGM. 
However,
it is not enough to understand the sources alone, 
we must also consider the sinks of ionizing radiation, 
i.e. the atoms that once ionized will recombine, wasting ionizing
photons that would otherwise contribute to reionization. 
Sinks will have two main effects on the 21-cm signal:
\begin{packed_enum}
\item as absorbers of ionizing photons they impact on both the
timing of reionization and the morphology of cosmic H II patches;
\item as reservoirs of neutral gas they contribute a 
non-zero 21-cm signal in otherwise ionized regions.
\end{packed_enum}

Sinks can reside in both the IGM and the interstellar medium (ISM), 
i.e. inside and outside of galaxies. 
In the ionized IGM at reionization redshifts, 
sinks mostly consist of diffuse structures and 
partially self-shielded gas clumps called Lyman-limit 
systems (LLSs, e.g. \citealt{Furlanetto2005}, \mbox{\citealt{McQuinn2011, Munoz2014}}).
LLSs are inhomogeneous in both space and time \citep{Crociani2011a, Choudhury2009};
as such, they can dramatically impact the duration and  morphology of reionization (effect i above; \citealt{Sobacchi2014}).
Although they are mostly ionized (e.g. \citealt{McQuinn2011, Rahmati2013}), 
LLSs can also contribute to effect (ii).
Indeed Sobacchi \& Mesinger (2014) recently estimated that LLSs on average contribute a few percent neutral fraction inside cosmic \hii patches during reionization.

Sinks are also present inside galaxies
where H\thinspace\textsc{i} can self shield
in dense clumps.  
At lower redshifts (e.g. \citealt{Wolfe2005} ),
sinks inside galaxies [mostly so-called Damped Lyman-alpha systems (DLAs); \citealt{Prochaska2005, Noterdaeme2012}] provide a larger reservoir of \hi compared to sinks in the IGM. Recent observations suggest that the contribution of DLAs
to the cosmological mass density is roughly constant until it 
increases from $z=2.3$ to $z=3.5$ by almost a factor 
of 2 \citep{Noterdaeme2012}.  There is currently no consensus in explaining these trends, making extrapolations to reionization redshifts uncertain.
The abundance and ionization structure of such galactic sinks are more susceptible to the local environment, including stellar feedback and baryonic cooling.  Therefore it is difficult to quantify the contribution of galactic sinks to effect (ii) above. Using a `tuning-knob' approach (similar to the one we use below), \citet{Wyithe2009} predict that by reducing the contrast between cosmic ionized and neutral patches (effect ii), galactic \hi could indeed damp the 21-cm power spectrum by 10-20\%.
 Galactic \hi also contributes to effect (i) by reducing the efficiency with which ionizing radiation escapes into the IGM, usually quantified with the so-called ionizing photon escape fraction, $f_{\rm esc}$. 
This clearly has an effect on the timing
of reionization.  However, reionization studies include the escape fraction and the associated uncertainties in the source term.

In this work we examine the effect that both LLSs (IGM sinks)
and differing levels of galactic H\thinspace\textsc{i} (galactic sinks)
have on the statistics of the 21-cm brightness temperature, 
with emphasis on 1-point statistics.
 We analyse the simulation presented in \citet{Sobacchi2014},
here after SM2014,
to isolate the impact of IGM sinks on the
1-point statistics via effects (i) and (ii). 
We also test how the statistics of this simulation
are altered in the presence of 
galactic sinks due to effect (ii).

The rest of this paper is structured as follows. 
In Section \ref{sec:sims} we describe the
simulations that we use to model reionization
including sinks in both the IGM and in galaxies;
in Section \ref{sec:analysis} we 
present our analysis methods and examine
the 21-cm maps and their probability-density 
functions; we then consider the evolution
of the 1-point statistics and our ability to
constrain them with radio telescopes present and future 
in Sections \ref{sec:noise} and
\ref{sec:smooth},
Section \ref{sec:ps} considers how much
the power spectrum might be altered in the
presence of galactic sinks and finally in 
Section \ref{sec:conc} we summarise our findings.
We use the following cosmological parameters ($\Omega_{\rm m}$,$\Omega_{\Lambda}$,$\Omega_{\rm b}$, 
$h$, $\sigma_8$, $n$) 
= (0.28, 0.72, 0.046, 0.70, 0.82, 0.96),
consistent with recent measurements by the Planck
satellite 
\citep{PlanckCollaboration2013b}.

\section{The simulations}\label{sec:sims}

To investigate the impact of residual \hi on 21-cm statistics, we make use of the public simulation tool, \cmfast\footnote{\url{http://homepage.sns.it/mesinger/Sim.html}}.  
The density and velocity fields in \cmfast\ are generated with standard perturbation theory \citep{Zel'dovichYa.B.1970}.
To generate ionization fields the excursion-set approach 
of \citet{Furlanetto2004} is applied to the evolved density fields: 
to determine whether a region is ionized, this algorithm compares 
at decreasing radii the number of ionizing photons to the number 
of baryons (plus recombinations).

\begin{figure}
  \centering
  \includegraphics{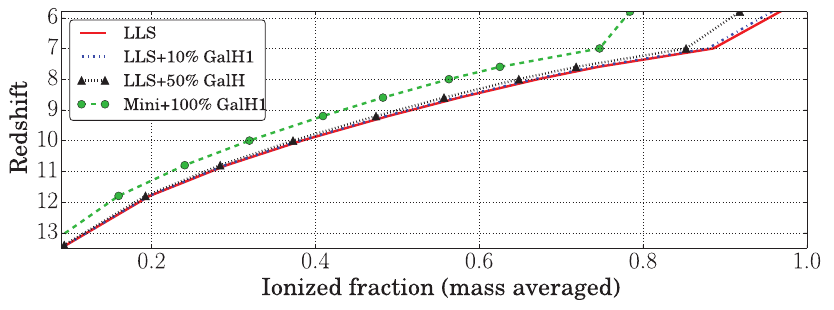} \\
  \caption{Evolution of mass-averaged ionized fraction for our 4 main models (by construction, the volume-averaged evolution is the same in all models). To track evolution of the morphology of H\thinspace\textsc{ii} regions, we plot these four models as a function of the H\thinspace\textsc{ii} volume-filling factor hereafter.}
  \label{fig:nfEvo}
\end{figure}

In this work, we study how sinks of ionizing photons impact the 21-cm signal, focusing on 1-point statistics.  To do this, we run four \cmfast\ realisations of the ionization field during reionization, varying the impact of \hi on both the large-scale reionization morphology and the level of residual \hi inside cosmic \hii regions (effects i and ii discussed in the introduction).  We discuss the details below, showing a summary of the four simulations used in Table \ref{tbl:model} and Figure \ref{fig:nfEvo}.
All simulations are $L=300$\,Mpc on a side with a resolution of 400$^3$.  The density and velocity fields are the same in all of the runs. 

\subsection{Homogeneous recombinations (\cmfastvone)}

We call our reference simulation \textbf{`NoLLS'}. Sinks have a minimal impact in this simulation.  It is generated with the current public release, \cmfastvone.  The resulting ionization fields are similar to ones obtained with radiative transfer simulations (e.g. \citealt{Zahn2011}).
In \cmfastvone, a region is ionized if its (time-integrated) number of ionizing photons exceeds the number baryons plus a global average number of recombinations, $\bar{n}_{\rm rec}$:
\begin{equation}
\label{eq:fcollv1}
\zeta f_{\rm coll}(\textbf{x}, z, R, M_{\rm min}) \geq 1+\bar{n}_{\rm rec} ~, 
\end{equation}
\noindent where $f_{\rm coll}(\textbf{x}, z, R, M_{\rm min})$ is
the fraction of matter, inside a region of scale $R$,
which resides in halos with mass greater than $M_{\rm min}$
(taken here to correspond to a virial temperature of 10$^4$ K).
The ionizing efficiency can be expanded as 
$\zeta = f_{\rm esc}f_\ast N_{\gamma/b}$,
where $f_{\rm esc}$ is the fraction of ionizing photons that 
escape galaxies, $f_\ast$ is the fraction of galactic baryons inside stars, and $N_{\gamma/b}$ is the number of ionizing photons produced per stellar baryon.
This ionization criterion in eq. (\ref{eq:fcollv1}) is checked in an excursion-set fashion \citep{Furlanetto2004}, starting from a maximum radius $R_{\rm max}$ 
\footnote{$R_{\rm max}$ is a free parameter based on ionized
photon mean free path at the redshifts of interest, see 
\citet{Storrie-Lombardi1994, MiraldaEscude2003, Choudhury2008}.
In our `NoLLS' run it is
set to
30\,Mpc.   However there is no maximum smoothing radius enforced by the
simulations of SM2014 since the ionized photon mean free path is determined 
by the properties of `LLS', which are computed self-consistently.}
down to the cell size.
At the last smoothing
scale, the unresolved, sub-cell \hii fraction is set to 
$\zeta (1+\bar{n}_{\rm rec})^{-1} f_{\rm coll}(\textbf{x}, z, R_{\rm cell}, M_{\rm min})$.

\begin{table}
    \caption{Model summary.}
    \begin{center}
      \begin{tabular}[p]{||l||l||l||}
	\hline
        \textbf{Model}  &  \textbf{Properties} &  \parbox{2.2cm}{\textbf{$\boldsymbol{\langle x_{\rm \textsc{hi}} \rangle_M}$ in H\thinspace\textsc{ii} regions at $z=9.20$, $\boldsymbol{ Q_{\rm \textsc{hii}} }\sim 0.5$ } } \\
        \hline
        \parbox{0.8cm}{NoLLS}
        &  \parbox{3.9cm}{ \cmfastvone; sinks are included as a constant in the ionizing efficiency.  Cosmic \hii patches are fully ionized.} 
        & 0.0\\ 
        \hline
        \parbox{0.8cm}{LLS}
        &  \parbox{3.9cm}{Self--consistent treatment of local 
          recombinations \& UVB feedback of SM2014. Small absorbers act as sinks 
          of ionizing radiation to slow the progress of reionization, 
          reduce the size of ionized regions 
          \& suppress the 21-cm temperature contrast 
          between H\thinspace\textsc{ii} (over-dense) and 
          H\thinspace\textsc{i} (under-dense) regions.         } 
        & \parbox{1.5cm}{0.03} \\ 
        \hline
        \parbox{0.8cm}{LLS + Gal H\thinspace\textsc{i}}
        &  \parbox{3.9cm}{`LLS' with the addition of a fixed percentage 
          of remnant H\thinspace\textsc{i} in galaxies; 
          its contribution to the 21-cm signal cosmic
          inside H\thinspace\textsc{ii}
          regions further suppresses the contrast 
          between ionized and neutral patches.} 
          & \parbox{1.8cm}{0.1$f_{\rm coll}$: 0.04  

          0.5$f_{\rm coll}$: 0.07}        \\ 
        \hline
        \parbox{0.9cm}{Mini + 100\% Gal H\thinspace\textsc{i}} 
        &  \parbox{3.9cm}{`LLS' with both sterile minihalos (cooled via H$_2$)
          as well as atomic cooling halos 100\%
          neutral. Describes the most 
          intense impact that remnant
          H\thinspace\textsc{i} could have on the 21-cm signal 
          by suppressing the contrast
          between cosmic ionized and neutral patches.
        }& \parbox{1.5cm}{0.19} \\
	\hline
      \end{tabular}
    \end{center}
    \label{tbl:model} 
\end{table}

The `NoLLS' run has a different reionization evolution and morphology from the other three runs.\footnote{The mean free path, $R_{\rm max}$, framework does not capture the sub-grid recombination physics of the SM2014 model, discussed in the next section.  Nevertheless, these authors find that a choice of $R_{\rm max} \sim 10$ Mpc results in a reionization morphology and evolution roughly consistent with their sub-grid model.  This `effective ionizing photon horizon' is in fact much smaller than the physical mean free path (defined based on the instantaneous recombination rate and emissivity), and is empirically derived.}  
In order to facilitate direct morphological comparisons, we vary 
$\boldsymbol{\bar{n}_{\mathrm{rec}}}$
in `NoLLS' to match the redshift evolution of the filling
factor of \hii regions, $Q_{\rm HII}$ from the other runs.

\subsection{Inhomogeneous recombinations in the IGM}

As the next step, our \textbf{`LLS'} run includes inhomogeneous recombinations in the IGM using the model of SM2014 (their `FULL' model).  Such recombinations, driven by $\sim$kpc scale gas clumps can slow the growth of large cosmic \hii patches, primarily resulting in reduction of large-scale reionization structure.  SM2014 implement these local recombinations by modifying the ionization criterion from eq. (\ref{eq:fcollv1}) to:
\begin{equation}
\label{eq:fcollv2}
\zeta f_{\rm coll}[\textbf{x}, z, R, \bar{M}_{\rm min}(\textbf{x}, z)]\geq 1+\bar{n}_{\rm rec}(\textbf{x}, z) ~.
\end{equation}
In the model of SM2014, both the number of recombinations and minimum halo mass hosting galaxies depend on the local properties of each $\sim$ 1Mpc-scale reionization cell.  Taking self-shielding into account via the parametrization of \citet{Rahmati2013}, and integrating over the entire density distribution, the recombination rate per baryon in an ionized cell is:
\begin{equation}
\label{eq:dn_rec}
\frac{d n_{\rm rec}}{dt}(\textbf{x}, z)=\int_0^{+\infty}P_{\rm V}\left(\Delta, z\right)\Delta\bar{n}_{\rm H}\alpha_{\rm B}\left[1-x_{\rm HI}\left(\Delta\right)\right]^{2}d\Delta ~,
\end{equation}
\noindent where $P_{\rm V}\left(\Delta, z\right)$ is the full (non-linear, sub-grid) distribution of overdensities: $\Delta\equiv n_H/\bar{n}_H$ \citep{Miralda-Escude2000}, and  $x_{\rm HI}\left(\Delta\right)$ is equilibrium neutral fraction computed using the empirical self-shielding prescription of \citet{Rahmati2013}.  Then the total, time-integrated number of recombinations per baryon, averaged over the local cosmic \hii patch can be written as:
\begin{equation}
\label{eq:n_rec}
\bar{n}_{\rm rec}(\textbf{x}, z)=\left\langle\int_{z_{\rm IN}}^{z}\frac{dn_{\rm rec}}{dt}\frac{dt}{dz}dz\right\rangle_{\rm HII}\,,
\end{equation}
\noindent where $z_{\rm IN}$ is the ionization redshift of a simulation cell. 
Photo-heating feedback from reionization (i.e. the depletion of the gas reservoir available for star-formation), is included in this simulation with a similar approach, in which the minimum halo mass capable of retaining gas is computed using the local (cell's) photo-ionization history \citep{Sobacchi2013}.
Photo-heating feedback has a sub-dominant impact on reionization morphology, compared with inhomogeneous recombinations; however, since the sources and sinks are spatially correlated, the effects are additive in that they both prolong reionization history.

`LLS' (corresponding to the `FULL' simulation of SM2014) is our fiducial model for reionization history and morphology.  The impact of inhomogeneous, sub-grid \hi on the large-scale reionization morphology can be estimated by comparing the `NoLLS' and `LLS' runs (see Figure \ref{fig:dTSliceCompare}).  Our `LLS' run also includes residual \hi inside the cosmic \hii patches, contributing on average a mass-averaged neutral fraction of a few percent inside \hii regions (SM2014; see table 1).

\subsection{Inhomogeneous recombinations in the IGM + galactic \hi}

Our remaining runs use the same large-scale reionization morphology and redshift evolution as `LLS' (see Figure \ref{fig:dTSliceCompare}).  However here we explicitly increase the level of residual \hi inside the ionized regions.  Since the amount of \hi inside galaxies is very difficult to predict, we adopt a `tuning-knob' approach similar to \citealt{Wyithe2009}.  Specifically, we compute the average\footnote{We ignore the Poisson noise associated with halo discreteness, since this affects scales too small to be observable with the SKA.} collapsed fraction inside each cell, using the conditional (dependent on the cell's density) mass fraction (e.g. \citealt{Bond1991,LaceyCedric1993}) normalized to the mean predicted by Sheth-Tormen (e.g. \citet{Jenkins2001}; see also the discussion in \citealt{Barkana2004, Signal2010}).
We then assign a fixed fraction, $\alpha$, of the cell's baryonic mass to be neutral, so that each cell's galactic \hi number density is $\alpha \Delta_{\rm cell} \bar{n}_H f_{\rm coll}(M_{\rm min}, z, \Delta_{\rm cell}, R_{\rm cell})$.  Keeping with the simplicity of the `tuning-knob' approach, we use a uniform value of $M_{\rm min}$ corresponding to a virial temperature of 10$^4$ K.
We emphasise that this implementation considers only the additional
signal that galactic sinks contribute; their impact
on reionization timing is assumed to be 
accounted for in the ionizing efficiency of sources.

We study 
different values of $\alpha$; for the sake of brevity,
we only present the simulations with 10\% and 50\% 
of the galactic mass as neutral hydrogen.
To provide a sense of the range of values
that $\alpha$ might take we can extrapolate lower
redshift constraints from DLAs. 
The contribution
of galactic neutral hydrogen to the critical 
mass is found to be $\Omega_{\textrm{\textsc{hi}}}\sim 10^{-3}$
for $z\lesssim 4$ \citep{Prochaska2005, Noterdaeme2012}.
This can be related to $\alpha$ as 
$\alpha = \Omega_{\textrm{\textsc{hi}}}/(\Omega_{\rm b}f_{\rm coll})$,
which predicts $\alpha\sim 0.05$ towards the
end of reionization, i.e $z \sim 6$,
rising to $\alpha\sim 1$ at $z > 11$ as $f_{\rm coll}$ decreases with 
increasing redshift. However at the redshifts
for which $\Omega_{\textrm{\textsc{hi}}}$ has been 
constrained, 
$\alpha\lesssim 0.03$;
as such, here after 
we will make the distinction
between `reasonable levels of galactic sinks', ($\alpha\lesssim 0.1$),
and `large levels' ($\alpha\gtrsim 0.5)$.

It is important to note that the tail of the \citet{Miralda-Escude2000} density distribution should, in principle, include gas inside galaxies.  However, the distribution was calibrated to the lower overdensity IGM, and has been shown to be a poor fit to the matter distribution inside lower-redshift galaxies, even neglecting feedback (e.g. \citet{Bolton2009, McQuinn2011}). 
Moreover, the accuracy of the \citet{Miralda-Escude2000} PDF is untested both at the high-redshifts ($z\sim10$) of interest here, as well as in predicting the conditional (dependent on the local, large-scale overdensity) mass fraction.  Hence in our approach we use the conditional, excursion-set approach, which is much better tested in predicting the collapse fraction.  However, since the excursion-set galactic \hi is added on top of the SM2014 neutral fraction field, our approach is bound to double count some galactic H\thinspace\textsc{i}, and the values of $\alpha$ should only be treated as approximate (lower limits).  We find that due to the inside-out nature of the epoch of reionization (i.e. biased \hii regions), the excursion-set approach generally predicts much higher collapsed fractions inside \hii regions than MHR00 (see e.g. table 1); thus we do not expect the “double-counting” to have a large impact on our conclusions.

\subsection{Inhomogeneous recombinations in the IGM + extreme galactic \hi}
Finally, we consider an extreme model with the maximal 
amount of residual neutral hydrogen, `Mini+100\% Gal H\thinspace\textsc{i}'. 
Following the same procedure as outlined in the previous section, we assign 100\% of the baryonic mass 
in $10^3 \leq T_{\rm vir}$ halos as H\thinspace\textsc{i}.
This model effectively assumes that
minihalos with a virial temperature of $10^3 \leq T_{\rm vir} < 10^4$ K,
i.e. those that have cooled by molecular cooling, 
have been sterilized by 
a Lyman-Werner background so that they cannot form stars (e.g. \citealt{Haiman2000a, Ricotti2001, Mesinger2006}).
Moreover, this model assumes that the time-scale of minihalo photo-evaporation
is much longer than the duration of reionization (e.g. \citealt{Shapiro2004,Iliev2005, Ciardi2006}).
Our  `Mini+100\% Gal \hi' run therefore serves to illustrate the most extreme impact that residual galactic \hi
 could have on 21-cm observations.

\section{Brightness-temperature maps \& moments} \label{sec:analysis}
\begin{table}
    \caption{Instrumental specifications assumed for noise calculations. 
LOFAR and SKA parameters are taken from \protect\citet{Mellema2013}, HERA 
from \protect\citet{Pober2013}/ private communications with A. Liu and MWA 
parameters from \protect\citet{Tingay2013}.}
    \begin{center}
      \begin{tabular}[p]{|l|l|l|l|l|}
	\hline
        \textbf{Parameter}  &  \textbf{MWA} &  \textbf{LOFAR} & \textbf{HERA} &  \textbf{SKA} \\
        \hline
        Number of stations \\
        ($\bmath{N_{\rm stat}}$)
        & 128 & 48 & 331 & 450\\ 
        \hline
        Effective area \\
        ($\bmath{A_{\rm eff}}$/m$^2$)
        & 21.5 & 804 & 8.4e3/$N_{\rm stat}$ & $10^6$/$N_{\rm stat}$\\ 
        \hline
        Maximum baseline \\
        ($\bmath{D_{\rm max}}$/m)
        & 2864 & 3000 & 360 & $10^4$\\ 
        \hline
        Integration time \\
        ($\boldsymbol{t_{\rm int}}$/hours) 
        &1000 & 1000 & 1000 & 1000 \\
        \hline
        Bandwidth \\
        ($\boldsymbol{B}$/MHz) 
        &6 & 6 & 6& 6 \\
	\hline
      \end{tabular}
    \end{center}
    \label{tbl:instStats} 
\end{table}
\begin{figure*}
\begin{minipage}{176mm}
\begin{tabular}{c}
  \includegraphics[scale=0.65]{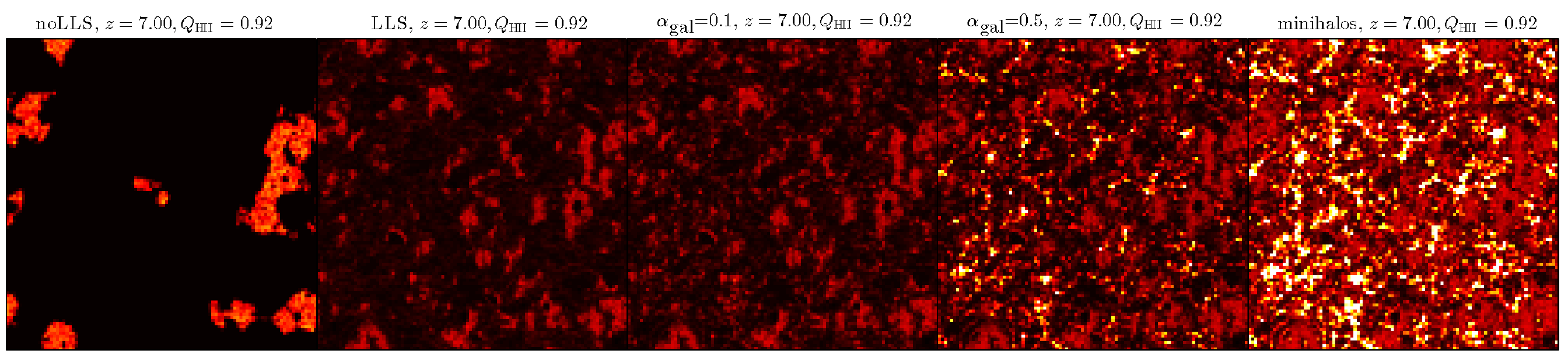} \\
  \includegraphics[scale=0.65]{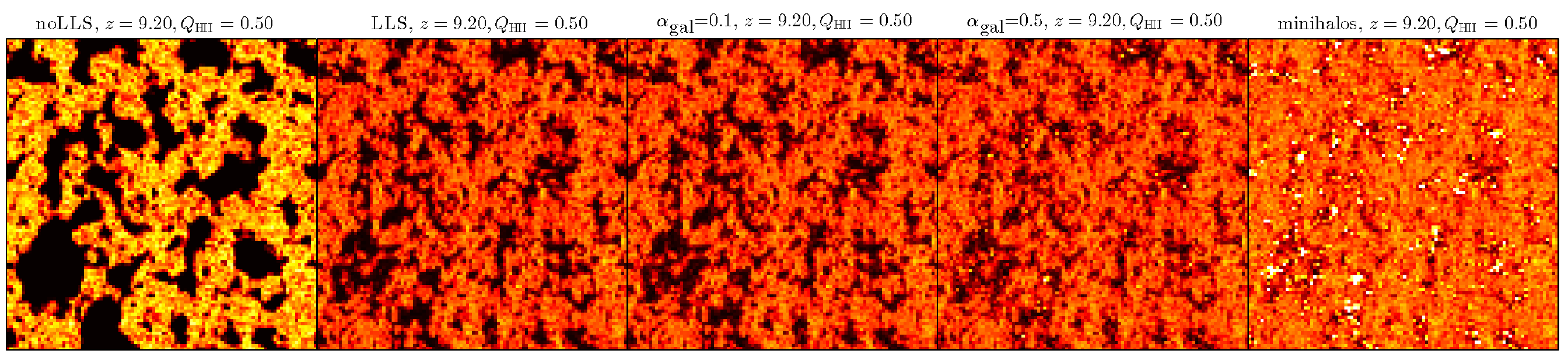} \\
  \includegraphics[scale=0.65]{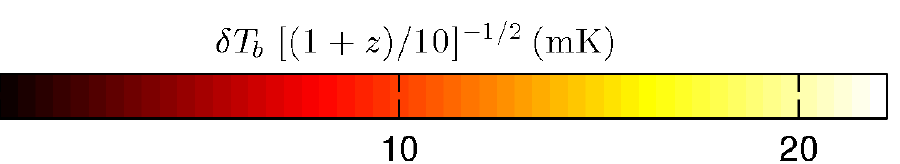} \\
\end{tabular}
\caption{Brightness-temperature maps for 300 Mpc boxes 
normalised to z = 9; maps are resolved 
to pixels of (3 Mpc)$^3$ and presented with the pixel 
depth flattened into the page.
Maps correspond to, from LLS (left), LLS + 10\%GalH\thinspace\textsc{i} (middle-left), LLS + 50\%GalH\thinspace\textsc{i} (middle-right) and 
minihalos + 100\%GalH\thinspace\textsc{i} (right).
The top row is for z=7.0 when the H\thinspace\textsc{ii}
volume filling factor is 0.92;
the bottom row corresponds to z=9.2 at which time the H\thinspace\textsc{ii}
volume filling factor is 0.50.  
Black regions are those
with zero brightness temperature.
}
\label{fig:dTSliceCompare}
\end{minipage}
\end{figure*}

\subsection{Computing 1D statistics and noise}
Once the ionization boxes have been calculated as 
described, the brightness temperature  ($\delta T_{\rm b}$) 
can be computed according to:
\begin{equation}
\begin{split}
\delta T_{\rm b}=&\frac{T_{\rm s}-T_{\gamma}}{1+z}(1-e^{-\tau_{\nu_0}})\\
\approx&\,27.\,\frac{T_{\rm s}-T_{\gamma}}{T_{\rm s}}\,x_{\rm \textsc{hi}}(1+\delta)\left[\frac{H(z)/(1+z)}{\d v_{\rm r}/\d r}\right]\\
&\times \left(\frac{1+z}{10}\frac{0.15}{\Omega_{\rm m}h^2}\right)^{1/2}\left(\frac{\Omega_{\rm b}h^2}{0.023}\right) \rm mK
\,,\\ \label{eqn:BrightTemp}
\end{split}
\end{equation}
\noindent where $\delta T_{\rm b}$ is dependent on the overdensity
 $\delta = \rho/\overline{\rho} -1$,  
velocity gradient $\d v_{\rm r}/\d r$
and cosmology. The gas spin temperature
 $T_{\rm s}$ measures the occupation levels 
of the two hyperfine energy levels involved 
in the H\thinspace\textsc{i} 21-cm transition 
and determines the 21-cm optical depth $\tau_{\nu_0}$.
It is assumed in this work that  
$T_{\rm s}\gg T_{\gamma}$, i.e. $T_{\rm s}$ is much 
higher than that of the cosmic microwave background $T_{\gamma}$.
Whilst this is expected to be a reasonable approximation 
during reionization due to X-ray heating, 
the influence of spin temperature fluctuations
may not be negligible in the early stages of reionization, 
i.e. $Q_{\rm H\textsc{ii}} < 0.2$ 
\citep{Pritchard2007, Mesinger2013a, Ghara2014}. 

In our analysis of the simulations described above
we use the same approach as \citet{Watkinson2014}, hereafter 
WP2014. 
We measure the variance $S_2$ and the skew $S_3$ of a simulated box 
according to
\begin{equation}
\begin{split}
S_2 &=  \frac{1}{N_{\rm pix}}\sum_{i=0}^{N_{\rm pix}}\left[\delta T_{i}-\delta\overline{T}_{\rm b}\right]^2\,,\\
S_3 &=  \frac{1}{N_{\rm pix}}\sum_{i=0}^{N_{\rm pix}}\left[\delta T_{i}-\delta\overline{T}_{\rm b}\right]^3\,,\\
\end{split}\label{eq:moments}
\end{equation}
\noindent where $\delta T_{i}$ is the brightness temperature
of the i$^{\rm th}$ pixel, the average brightness temperature
in a box is given by
$\delta \overline{T}_{\rm b} =N_{\rm pix}^{-1}\sum_{i=0}^{N_{\rm pix}} \delta T_{i}$
and $N_{\rm pix}$ is the total number of cells in a box. 
We will consider two normalisations for the skew, the 
standard skewness $S_3/(S_2)^{3/2}$ and the dimensional
skewness $S_3/S_2$ which was found to be a more natural 
normalisation for the skew of the neutral-fraction boxes of WP2014.

We make order of magnitude calculations for the
error induced by instrumental noise (using the instrument parameters in  table \ref{tbl:instStats}), but do not consider 
foreground residuals.  We assume the noise is
independent on each pixel but identically distributed
in a manner well described by Gaussian random noise with 
zero mean and standard deviation $\sigma^2_{\rm noise}$ where
\begin{equation}
\begin{split}
\sigma^2_{\rm noise}= 2.9 \rm{mK}&\left(\frac{10^5\rm{m}^2}{A_{\rm tot}}
                \right)\left(\frac{10'}{\Delta\theta}
                   \right)^2\\
                   &\times \left(\frac{1+z}{10.0}
                          \right)^{4.6}
                          \sqrt{\left(\frac{1\rm{MHz}}{\Delta\nu}\frac{100\rm{hours}}{t_{\rm int}}
                                     \right)} \,.
\\
\label{eq:InstNoise2}
\end{split}
\end{equation}
This expression is derived with the noise 
on the brightness temperature described by
$ \Delta T^N = T_{\rm sys}/\eta_{\rm f}\sqrt{\Delta\nu t_{\rm int}}$
where the array filling factor is defined as 
$\eta_{\rm f}=A_{\rm tot}/D^2_{\rm max}$ in which $A_{\rm tot}$ is 
the total effective area of the array, 
$D_{\rm max}=\lambda/\Delta\theta$; the system
temperature $T_{\rm sys}$ is assumed to be 
saturated by the sky temperature at the 
frequencies of interest. 
Before measuring the moments we smooth and resample the boxes 
to correspond
roughly with the resolution expected from the various instruments; 
to do so, we use a Gaussian smoothing kernel with width
$R_{\rm pix}=6$\,Mpc for LOFAR/ MWA and $R_{\rm pix}=3$\,Mpc for 
HERA/ SKA (although we note that in this work, 
LOFAR/ MWA results are only presented for larger smoothing scales). 
We also reduce the assumed frequency resolution
of each telescope to correspond to $R_{\rm pix}$, i.e.
$\Delta\nu=H_0\nu_0\sqrt{\Omega_{\rm m}}R_{\rm pix}/[c \sqrt{(1+z)}]$.

To estimate how the noise we've so far discussed will 
propagate onto the skewness and variance we assume
that each pixel has a measured signal $x_i=\delta T_{i} + n_i$,
where the noise on the pixel $n_i$ 
obeys the properties discussed above. 
We then 
use a test statistic for the $m^{\rm th}$ moments, 
$N_{\rm pix}^{-1}\sum_{i=0}^{N_{\rm pix}}(x_i-\overline{x}_i)^m$,
to determine unbiased estimators for the moments
of Equation \ref{eq:moments}. 
We then calculate the noise-induced variance on each unbiased estimator.
These derived 1-$\sigma$ errors are then propagated onto
the normalised skewness estimators  
$\gamma_3=\hat{S}_3/\hat{S}_2$ and $\gamma'_3=\hat{S}_3/(\hat{S}_2)^{3/2}$
to give
\begin{equation}
\begin{split}
V_{\hat{S}_2}&= 
\frac{2}{N} (2S_2\sigma_{\rm noise}^2 
+ \sigma_{\rm noise}^4)\,,\\
V_{\gamma_3} &\approx \frac{1}{(S_2)^2}V_{\hat{S}_3}
+\frac{(S_3)^2}{(S_2)^4}V_{\hat{S}_2}
-2\frac{S_3}{(S_2)^3}C_{S_2S_3} \,,\\
V_{\gamma'_3} &\approx \frac{1}{(S_2)^3}V_{\hat{S}_3}
+\frac{9}{4}\frac{(S_3)^2}{(S_2)^5}V_{\hat{S}_2}
-3\frac{S_3}{S_2^{4}}C_{\hat{S}_2\hat{S}_3} \,,\\
\end{split}
\end{equation}

\noindent where

\begin{equation}
\begin{split}
V_{\hat{S}_3}&= 
\frac{3}{N_{\rm pix}}(3\sigma_{\rm noise}^2K_4
+12S_2\sigma_{\rm noise}^4
+5\sigma_{\rm noise}^6)\,, 
\end{split}
\end{equation}
\noindent and
\begin{equation}
\begin{split}
C_{\hat{S}_2\hat{S}_3}
&= \frac{6}{N_{\rm pix}} 
S_3 \sigma^2_{\rm noise} 
\,.\\
\end{split}
\end{equation}

\subsection{Brightness-temperature maps}

It is useful to take a look at the maps themselves
before analysing their statistics. In the maps 
of Figure \ref{fig:dTSliceCompare},
`NoLLS', `LLS', `LLS + 10\%GalH\thinspace\textsc{i}', `LLS + 50\%GalH\thinspace\textsc{i}' and 
`minihalos + 100\%GalH\thinspace\textsc{i}' 
are represented from left to right; 
the top strip corresponds to z = 7.0
and $Q_{\mathrm{\textsc{hii}}} \sim \boldsymbol{0.92}$
and the bottom strip corresponds to z = 9.20
and $Q_{\mathrm{\textsc{hii}}} \sim 0.50$. 
All maps have been smoothed and resampled to produce pixels
of 3 Mpc on a side, the pixel depth has been 
flattened into the page.

As mentioned above, the impact of inhomogeneous IGM sinks can be seen by comparing the large-scale reionization morphology of the left panel to the other panels.  As discussed in SM2014, recombinations in unresolved structures result in a dramatic suppression of large-scale \hii regions.  Comparing at a fixed $Q_{\rm HII}$, the distribution of \hii patch sizes is both narrower and shifted to smaller scales.  The impact is dramatic enough that partially ionized cells, hosting unresolved \hii bubbles, become important.  The ionized fraction of these cells is only $\lsim$10\%, but since they correspond to the densest (of those not yet reionized) $\sim$Mpc-scale patches, this is sufficient to damp the signal from the neutral cosmic patches (as evidenced by the `yellow' vs `orange' IGM surrounding the \hii bubbles in the `NoLLS' and `LLS' panels). This damps the contrast between the fully ionized and fully neutral regions. We caution however that unresolved, sub-grid \hii regions are taken into account only in a simplistic fashion in our model, as described above; further work is required to accurately model such sub-grid physics.

The impact of an increasing level of residual \hi inside the cosmic \hii patches can be seen by comparing the insides of the `black' regions in the panels of the Figure \ref{fig:dTSliceCompare} from left to right.  The model of SM2014 (our `LLS') only results in a few percent level of \hi inside \hii regions, and so on this colour scheme the \hii regions are still black.  However the extreme galactic \hi models, `LLS + 50\%GalH\thinspace\textsc{i}' and 
`minihalos + 100\%GalH\thinspace\textsc{i}', show a notable level of \hi inside the cosmic \hii regions, with small-scale structure tracing the distribution of galaxies.  This further reduces the contrast between neutral and ionized regions, complicating an interferometric detection of reionization.  In principle however, resolving this small-scale \hi structure with high-resolution 21-cm interferometry could constrain models of galactic \hi.

\subsection{Probability denstity functions}

\begin{figure}
  \centering
  \includegraphics{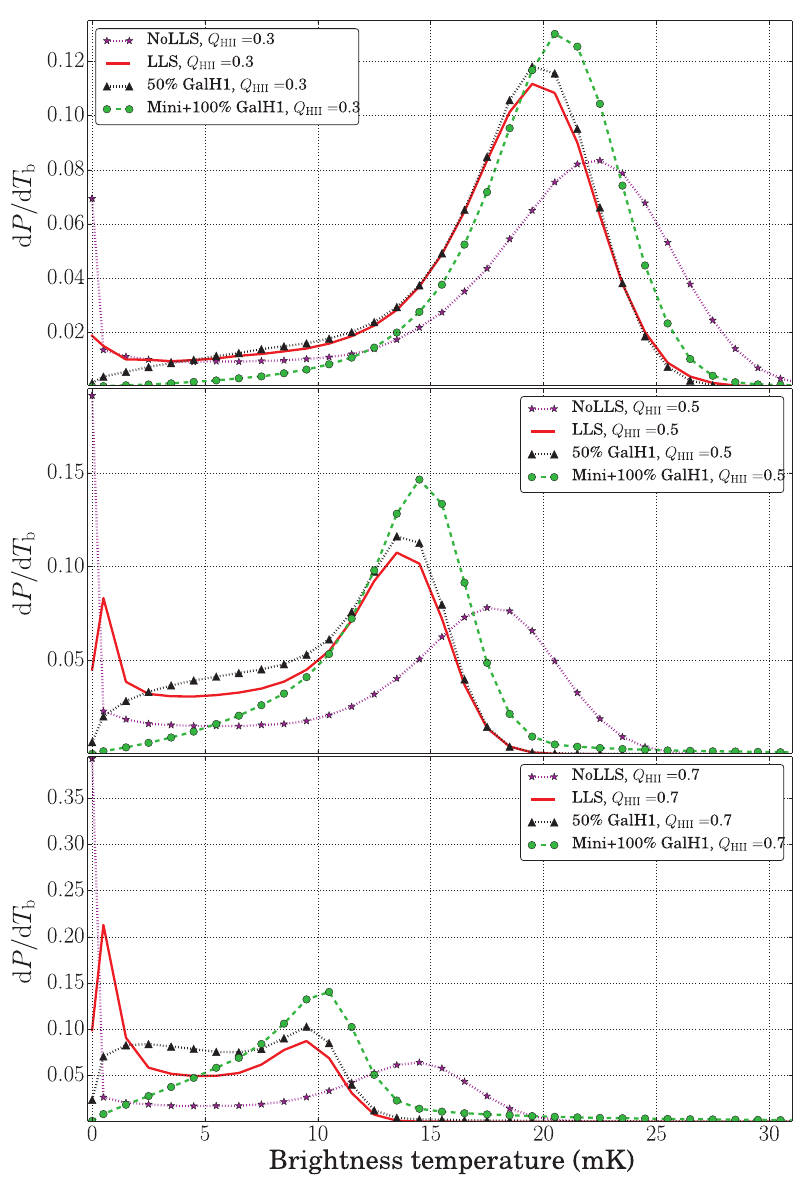} \\
  \caption{Brightness temperature PDFs at 
$Q_{\rm HII} \approx$ 0.3, 0.5, 0.7 ($z=$10.8, 9.2, 8), 
top to bottom. 
All maps are smoothed to a co-moving pixel size of 3 Mpc. 
}
  \label{fig:PDF}
\end{figure}

These trends are quantified in Figure \ref{fig:PDF}, where we present the probability density functions (PDFs) from these maps,
with lines corresponding to: `NoLLS'
(pink dashed w/stars),
`LLS' (red solid), `LLS + 50\% galactic H\thinspace\textsc{i}' (black dotted w/triangles)
and `minihalos + 100\% galactic H\thinspace\textsc{i}' (green dashed w/circles).
Plots correspond to $Q_{\rm HII} \approx$ 0.3, 0.5, 0.7 ($z=$10.8, 9.2, 8), top to bottom.

The presence of residual \hi inside cosmic \hii regions 
 suppresses zero-valued pixels in the PDFs.
The bi-modal nature of the PDF is 
maintained for `LLS' but the
sharp peak at $\delta T_{\rm b}=0$ mK
is smeared into a wider peak centred on
$\delta T_{\rm b}\sim 1$mK because of a few percent 
level of residual H\thinspace\textsc{i} in the ionized IGM (SM2014).
It is only with 
large levels of galactic sinks 
that the bi-modal nature is entirely suppressed,

Also evident in these PDFs is the impact
of sub-cell, unresolved \hii regions; these 
preferentially occur in the highest 
density pixels hosting newly forming galaxies, which is why there is
a suppression of the high $\delta T_{\rm b}$
tail in `LLS' when compared to `noLLS'.
This effect conspires with the 
loss of a sharp zero peak to further reduce the
variance in the maps.

\subsection{Observing the brightness-temperature moments: Instrumental noise}\label{sec:noise}
\begin{figure}
  \centering
  \includegraphics{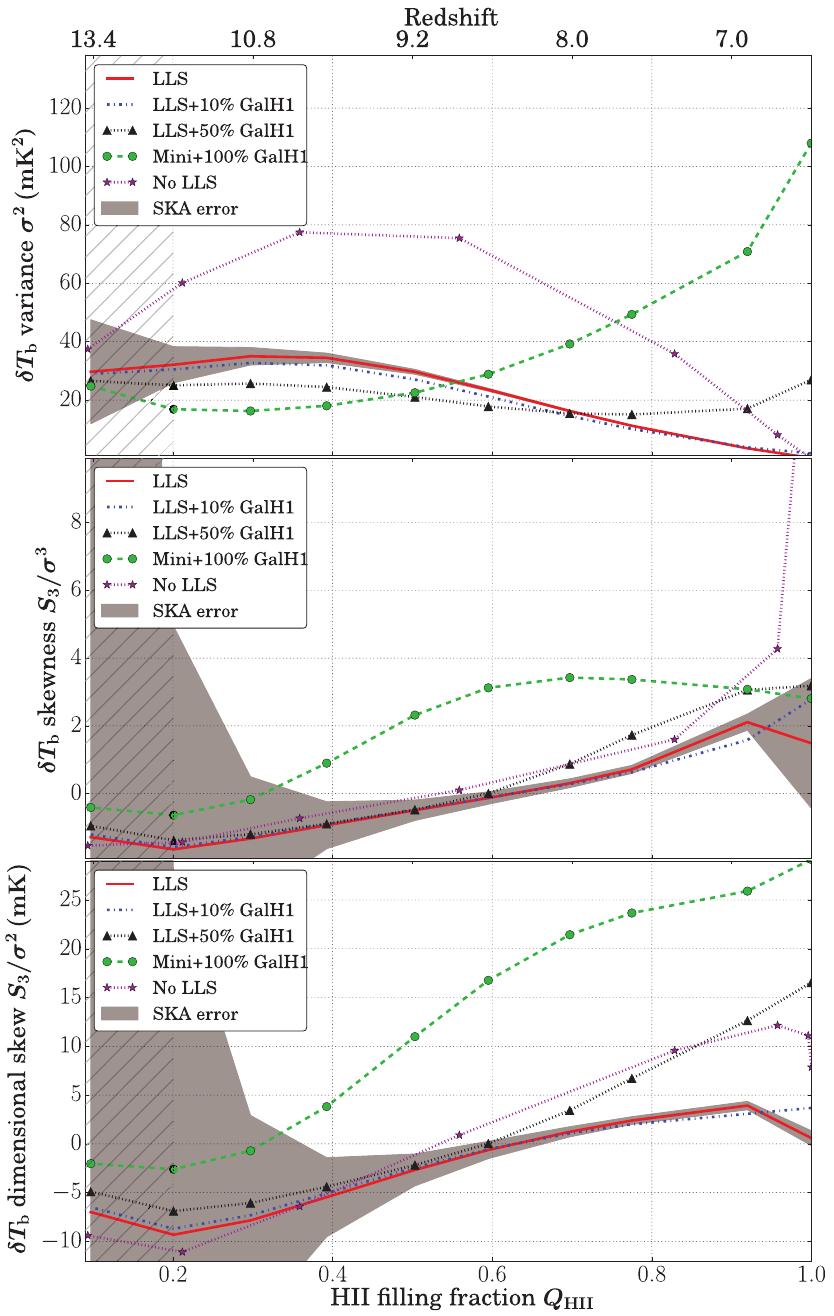} \\
  \caption{Variance (top), skewness (middle) and dimensional
skewness (bottom) of brightness temperature measured from
`observed' maps with a co-moving pixel size of 
 3 Mpc as a function of volume-filling factor of
ionized regions. 
Beige shadings depict SKA's 1-$\sigma$ instrumental errors 
for the `LLS' simulation.
Within the hatched region, 
spin-temperature fluctuations may not be negligible.
}
  \label{fig:dTSKAVariance}
\end{figure}

Figure \ref{fig:dTSKAVariance} shows evolution of the
variance (top), 
skewness (middle) and dimensional
skewness (bottom) as a function of the 
H\thinspace\textsc{ii} volume-filling factor.
Lines correspond to `LLS' (red solid),  `LLS + 10\% galactic H\thinspace\textsc{i}' (blue dot-dashed), `LLS + 50\% galactic H\thinspace\textsc{i}' (black dotted w/triangles), `minihalos + 100\% galactic H\thinspace\textsc{i}' (green dashed w/circles) and `NoLLS' (pink dotted w/ stars); we will use this key for the remains of the paper.

We begin by reviewing
what simulations that ignore recombinations in unresolved systems
predict for the evolution of 
the brightness-temperature moments.
Studies of statistics from such models identified several important 
properties that would provide invaluable constraints
on the timing of reionization if borne out in reality
(e.g. \citealt{Furlanetto2004, Harker2009, Watkinson2014}).
It is useful for the discussion that follows 
to refer to the pink dotted lines w/stars
in Figure \ref{fig:dTSKAVariance}.
The amplitude of the variance is maximised at the half-way mark
of reionization as the weight of the non-zero distribution of the
brightness-temperature's PDF and its spike at
zero are balanced (Figure \ref{fig:dTSKAVariance} top).
The late-time dominance of the zero-temperature spike induces a
rapid increase in skewness at late times (Figure \ref{fig:dTSKAVariance} middle). 
It could be argued that 
this effect is driven by the variance becoming very small as 
reionization draws to a close, however the dimensional skewness 
still exhibits a late-time signature but in the form of a turnover
 (Figure \ref{fig:dTSKAVariance} bottom).
At the halfway point both skewness statistics pass from 
negative to positive.
At early times they exhibit a minimum as the 
PDF transitions from being dominated by the density field 
to being dominated by the neutral fraction \citep{Lidz2008}; 
the location of this minimum depends on the overlap of X-ray heating and reionization epochs \citep{Mesinger2013a}.

Unresolved \hi sinks can dramatically change this picture.
Already with `LLS' the magnitude of the variance is suppressed by
up to a factor of two during the mid phases of reionization. 
Here the additional signal in H\thinspace\textsc{ii}
regions, reduces the contrast between
ionized and neutral regions.  More importantly, 
smaller H\thinspace\textsc{ii} regions
mean that there is more sub-pixel, unresolved ionization structure in `LLS' compared with `NoLLS'.  This suppression of bi-modality (on resolvable scales) in the ionization structure is further exacerbated by the more disjoint \hii regions in `LLS' which get `smeared-out' when the maps are smoothed; i.e. the \hii regions in `LLS' are less likely to be resolved by 21-cm interferometers.  These effects dramatically suppress the small-scale contrast between ionized and neutral regions; the strength of the turnover  associated with the halfway point of reionization is suppressed to the point that even SKA could have difficulty in constraining it. 
Fortunately, as we will see in the next section, 
further smoothing of the maps emphasises the turn-over feature,
improving our ability to constrain this mid-phase signature.
The turn-over in the variance 
occurs at $Q_{\mathrm{ \textsc{hii} }}\sim$ 0.4 in `LLS' rather than
nearly exactly halfway through reionization as predicted
by `NoLLS'. This is because the ionization map is
no longer well described as a two-phase field
in which the number of pixels in each 
phase perfectly balance
at the mid point of reionization.

The inclusion of galactic \hi reduces the variance even further (see also \citealt{Wyithe2009}). Reasonable levels of 
galactic sinks 
do not dramatically alter the evolution of the variance (as could already be seen from the maps in Figure \ref{fig:dTSliceCompare}).  However, if the galactic \hi is tied to a fixed fraction of $f_{\rm coll}$ as in our simple illustrative models, the growth of collapsed structure will boost the 21-cm variance at late times.
In the extreme minihalo+100\%Gal H\thinspace\textsc{i}
model, structure growth totally dominates
the signal throughout.
SKA could easily detect such a feature.

We see in the
middle panel of Figure \ref{fig:dTSKAVariance} 
that the transition from negative to positive
skewness including sinks occurs at $Q_{\mathrm{\textsc{hii}}}\sim 0.6$ (instead of $Q_{\mathrm{\textsc{hii}}}\sim 0.5$ as predicted by `NoLLS').  This transition is reduced to lower $Q_{\mathrm{\textsc{hii}}}$
in the presence of extreme levels of galactic sinks. 
For reasonable levels of galactic sinks, 
this is predominantly due to effect (i), 
where 
small H\thinspace\textsc{ii} regions
are unresolved. We even see this
trend to a lesser degree when smoothing of `NoLLS' 
on large scales smears-out the  
H\thinspace\textsc{ii} regions 
(see Figure \ref{fig:MomentsR10}).
For reasonable levels of galactic sinks,
this signature and the turn-over in the variance bracket the
mid-point and will constrain the timing of reionization
more tightly than either statistic on its own.

The late-time rapid increase 
predicted for the skewness is
totally suppressed by the presence of IGM sinks; 
instead we see a turn over in the skewness of 
`LLS' as the H\thinspace\textsc{ii}
filling factor reaches about 90\%. 
This is because the normalising factor, i.e.
the variance, is no longer tending towards zero 
and we instead see a transition from a skewness
driven by the ionization field to
one driven by the distribution of IGM sinks.
This turn over is wiped out by
the presence of even small amounts of  
galactic sinks.
These are important points to 
be aware of as we begin our efforts to constrain
reionization with 21-cm observations. 
Detecting a turnover in the skewness would 
imply the presence of IGM sinks 
(as would reduced variance) and
that reionization is in its final phases. The 
non detection of such a late-time signature in the 
skewness would not necessarily mean that we are not 
observing the end of reionization.
Its absence combined with evolution in the variance with
redshift would instead imply the presence of galactic sinks. 

SKA will be sensitive to the details of
skewness we have described above, but only in the dimensional
skewness, which exhibits more distinct features 
and better robustness to noise.

The early-time minimum in the skewness statistics is seen to 
be robust across the models.  This is understandable, since the location of this feature is determined by the overlap of X-ray heating and reionization \citep{Mesinger2013a}, and here we assume $T_S \gg T_\gamma$ throughout. Nevertheless,
extreme levels of residual H\thinspace\textsc{i}
will almost totally suppress this signature.
At its default resolution SKA could struggle to 
detect this feature, depending on the timings of
reionization; however, its non-detection would still provide
upper limits on this phase.

\begin{figure}
  \centering
  \includegraphics{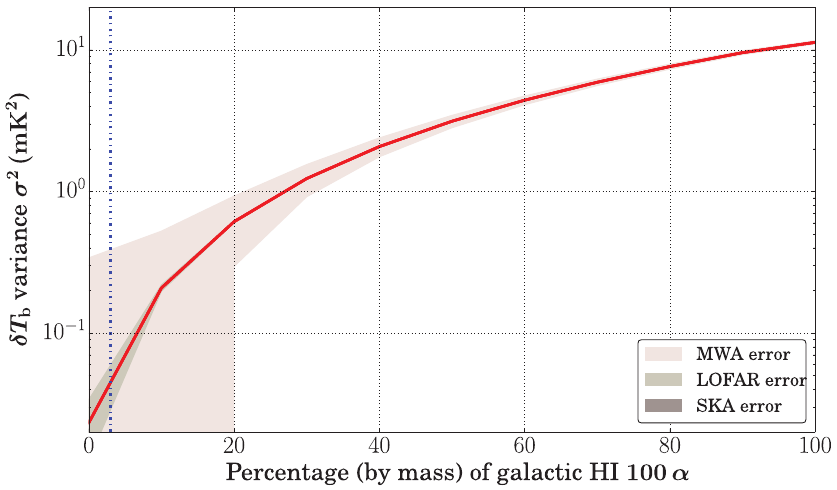} \\
  \caption{Variance as a function of percentage 
of remnant galactic H\thinspace\textsc{i} at z=5.80 ($Q_{\rm HII}=1$)
measured in `observed' maps smoothed to a
radius of $R_{\rm smooth}= 10$ Mpc to reduce noise,
the signal is around a factor of 10
greater with the default pixel size of 3 Mpc.
The vertical line corresponds to, $\alpha = 0.03$, 
illustrating the level
of residual hydrogen observed in galaxies at lower redshifts.
MWA, LOFAR and SKA 1-$\sigma$ 
instrumental errors are depicted with beige shadings
from lightest to darkest respectively.}
  \label{fig:ResGalHI}
\end{figure}

Figure \ref{fig:ResGalHI} shows the variance
as a function of the percentage of 
galactic mass in sinks
at the end of reionization, for this simulation $z=5.8$.
To reduce noise the maps are smoothed to a radius of 
$R_{\rm smooth}= 10$ Mpc and the beige shadings 
correspond to, from
light to dark, the MWA, LOFAR and SKA 
1-$\sigma$ instrumental errors.
Note that the signal when measured with SKA-like
resolution, i.e. 3 Mpc pixels, is greater by around a factor of ten.
The blue dot-dashed line marks the level of
residual hydrogen as constrained by DLAs at
lower redshift, i.e. $\alpha = 0.03$ at $z\sim 4$.

There are two ways to view this post-reionization
signal, 
one is that it is intrinsic noise to be overcome in 
our quest to constrain reionization, the other
is that it provides a constraint on 
residual H\thinspace\textsc{i} post-reionization.
Clearly the amplitude of the variance at the end of reionization 
is very sensitive to the level of galactic 
sinks. 
This intrinsic noise will be
detectable by SKA, which will therefore be well placed to 
constrain galactic sinks 
with this statistic. 
We do not explicitly
plot HERA errors but they are indistinguishable
from SKA, i.e. negligible, on this plot and
so should perform equally well at this task. 
Before these telescopes come
online, both LOFAR and MWA should be able to set some
constraints on this quantity, although at the lower
and more probable values of $\alpha$, MWA's signal to
noise will be too poor to do more than set upper
limits on remnant galactic H\thinspace\textsc{i}. 
The flip-side of this is that the intrinsic noise
from 
galactic sinks should not be
a limiting factor for these first generation instruments.

\subsection{Observing the brightness-temperature moments: Smoothing to reduce noise} \label{sec:smooth}

\begin{figure}
  \centering
  \includegraphics{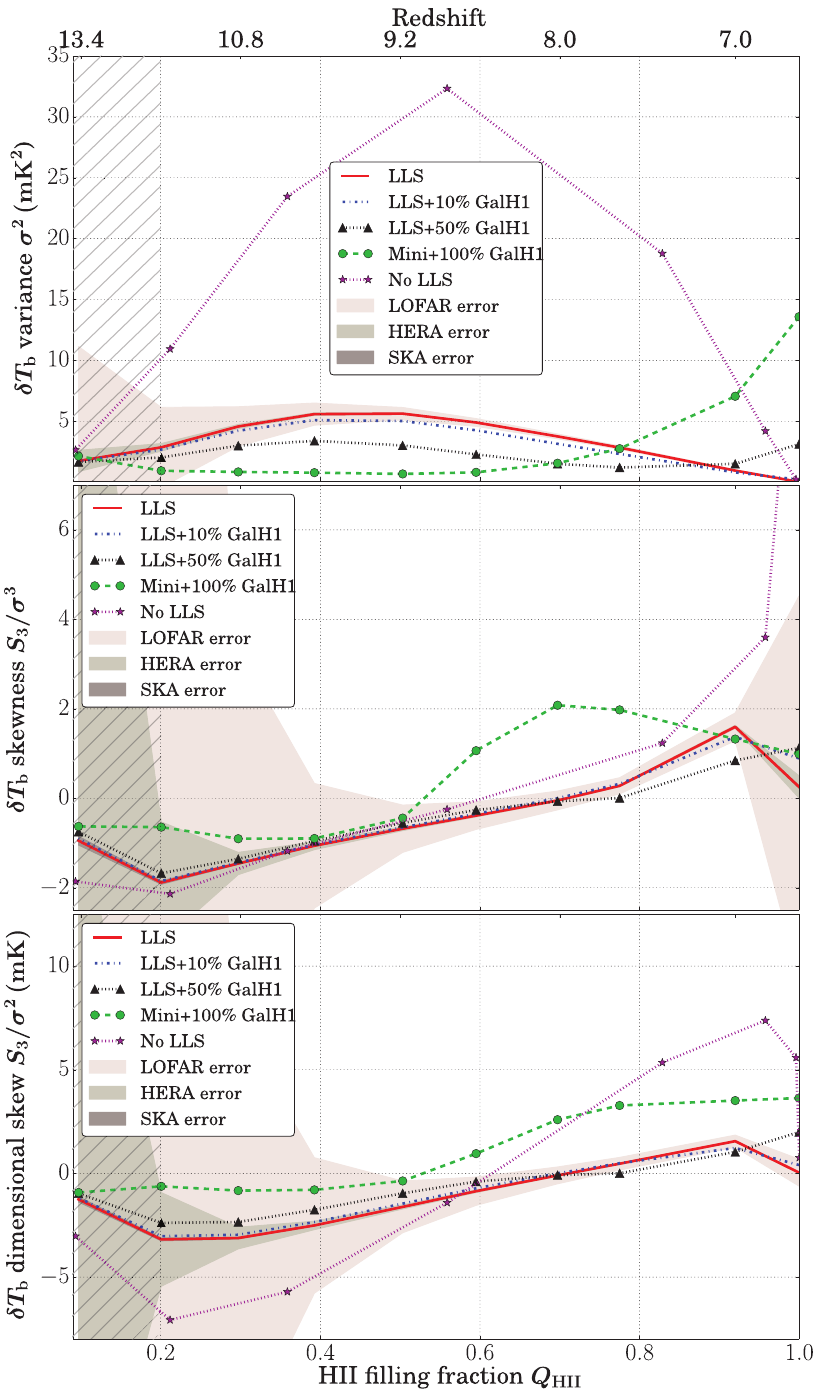} \\
  \caption{Variance (top), skewness (middle) and 
dimensional skewness (bottom) of brightness temperature 
as a function of the H\thinspace\textsc{ii}
volume-filling factor
measured in `observed' maps smoothed to $R_{\rm smooth}=$10\,Mpc. 
Beige shadings depict 1-$\sigma$ 
instrumental errors 
for the `LLS' simulation; tones correspond to MWA,
LOFAR and SKA from lightest to darkest.
Within the hatched region,
spin-temperature fluctuations may not be negligible.
}
  \label{fig:MomentsR10}
\end{figure}

As in \citet{Watkinson2014}, we investigate the effects of
smoothing the maps to beat down instrumental noise.
Figure \ref{fig:MomentsR10} (top) shows the evolution of the
variance when smoothed to $R_{\rm smooth}=$10\,Mpc as a function
of H\thinspace\textsc{ii} volume-filling factor. 
LOFAR will be well placed
to constrain the presence of IGM sinks through the reduction
in the variance, 
could constrain/ exclude 
large levels of galactic sinks  
and will
be able to offer timing constraints from the mid-point turnover 
in the variance
(unless the model is extreme and this signature does not exist).
We see that smoothing recovers
the late-time turnover in both skewness statistics
for reasonable levels of galactic sinks.
This is fortunate as the late-time maximum in the dimensional skewness 
will be detectable even by LOFAR for a wide range of 
galactic H\thinspace\textsc{i} 
levels. 
The mid-point of reionization will be bounded
by SKA and HERA measurements of the 
skewness, which will inform us when $Q_{\rm \textsc{hii}}<0.7$, 
and constraints from the variance of $Q_{\rm \textsc{hii}}\approx 0.4$.
The early-time minimum should still be detectable
in both skewness statistics despite suppression by the
presence of IGM sinks.
SKA is the only instrument that is likely
to actually detect this feature; however LOFAR and HERA should
be able to use its absence to set upper-limits
on this phase.
If residual hydrogen levels are extreme then 
this signature, like every other, will 
be suppressed.
 
\section{Brightness-temperature power spectrum} \label{sec:ps}
The brightness temperature's 1-point statistics form the focus of this 
paper, but the 21-cm power spectrum will be an incredibly
important statistic when it comes to constraining reionization.
As such it is worth understanding the impact of 
galactic neutral hydrogen on this statistic. 
We concentrate on the spherically-averaged 
dimensionless power spectrum which we describe using 
$\Delta^2_{\delta T_{\rm b}}(k,z)=k^3/(2\pi^2V)\langle |\delta_{21}(\bmath{k},z)|^2\rangle_k$ in which $\delta_{21}=\delta T_{\rm b}(\bmath{k},z)/\delta \overline{T}_{\rm b}(z)-1$, $\delta \overline{T}_{\rm b}(z)$ is the redshift dependent 
average brightness temperature calculated from the simulation, 
$V$ is the volume of the simulated
box and the angle brackets denote an average over $k$-space.
To model instrumental errors on 
the spherically averaged power spectrum, we take the 
approach outlined in the appendix 
of \citet{McQuinn2006} adopting a logarithmic bin width of
$\epsilon=0.5$. In calculating the number
density of baselines we assume a 
filled nucleus limited by the minimum
separation of stations, $\sqrt{A_{\rm eff}}$, 
followed by an $r^{-2}$ drop off in station 
density. 
This assumes a smooth density of 
stations, which for LOFAR is
a particularly crude approximation.
Whilst the exact configuration 
of SKA is yet to be decided,
both LOFAR and MWA have 
optimized station positioning,
as such our power spectrum errors are
indicative only. 

\begin{figure}
  \centering
  \includegraphics{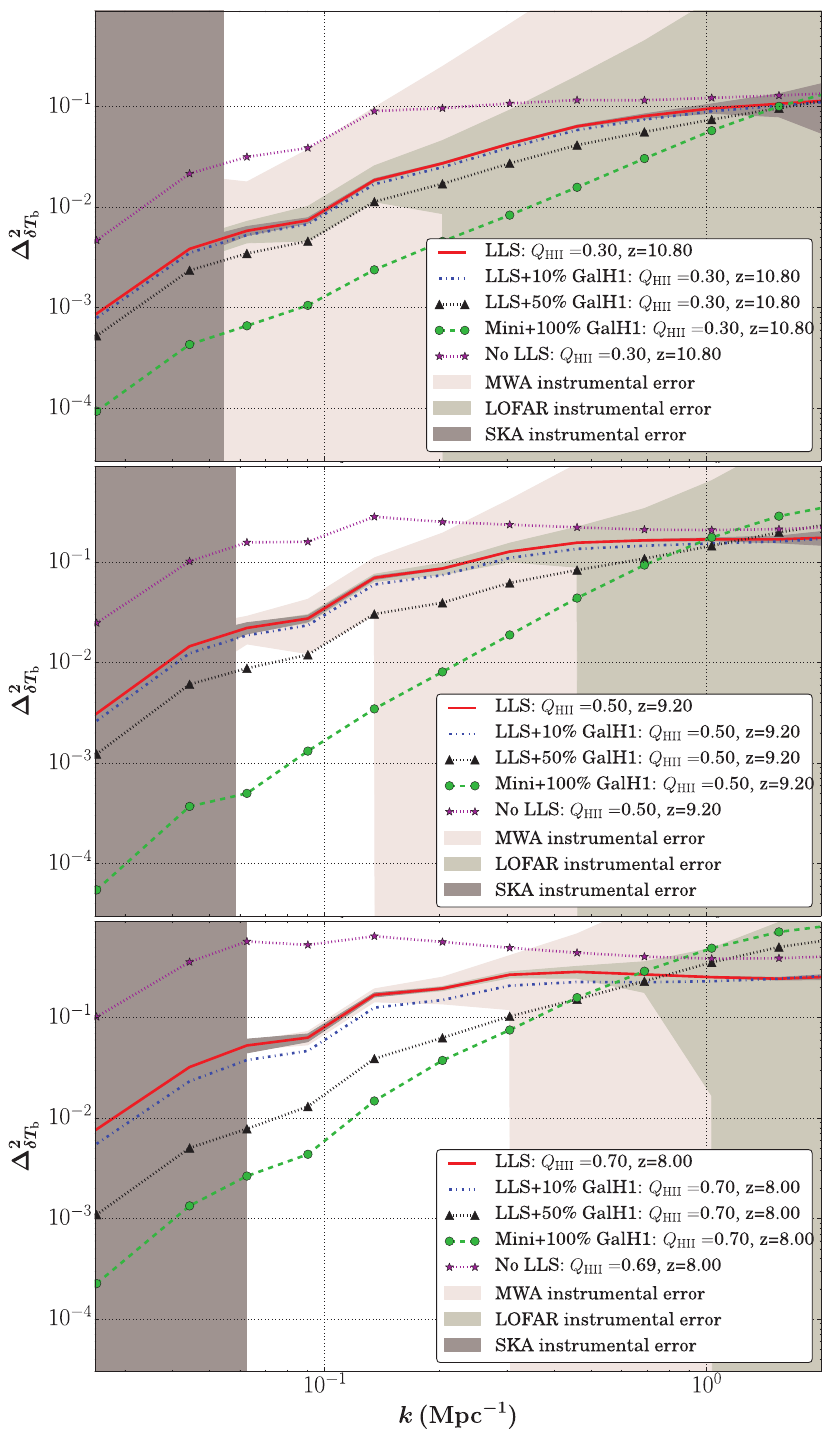} \\
  \caption{Dimensionless power spectrum of the brightness temperature
as a function of $k\,$Mpc$^{-1}$ for $z=10.8$ (top), $z=9.2$ (middle), 
and $z=8$ (bottom).}
  \label{fig:PS_v_k}
\end{figure}

The dimensionless power spectra for our simulations are presented in Figure \ref{fig:PS_v_k} for $Q_{\rm HII} \approx$ 0.3, 0.5, 0.7 ($z=$10.8, 9.2, 8), from top to bottom respectively.
As already discussed in SM2014, IGM sinks can suppress large-scale 21-cm power, resulting in much steeper power spectra throughout reionization.
From Figure \ref{fig:PS_v_k} we see that adding additional galactic \hi does not have a large impact on the power spectrum, for reasonable values of $\alpha$.  More extreme levels of galactic \hi result in a further steepening of the power spectrum, since: (i) the contrast between ionized and neutral patches is reduced, and (ii) the distribution of galactic \hi contributes additional power on small-scales.

\begin{figure}
  \centering
  \includegraphics{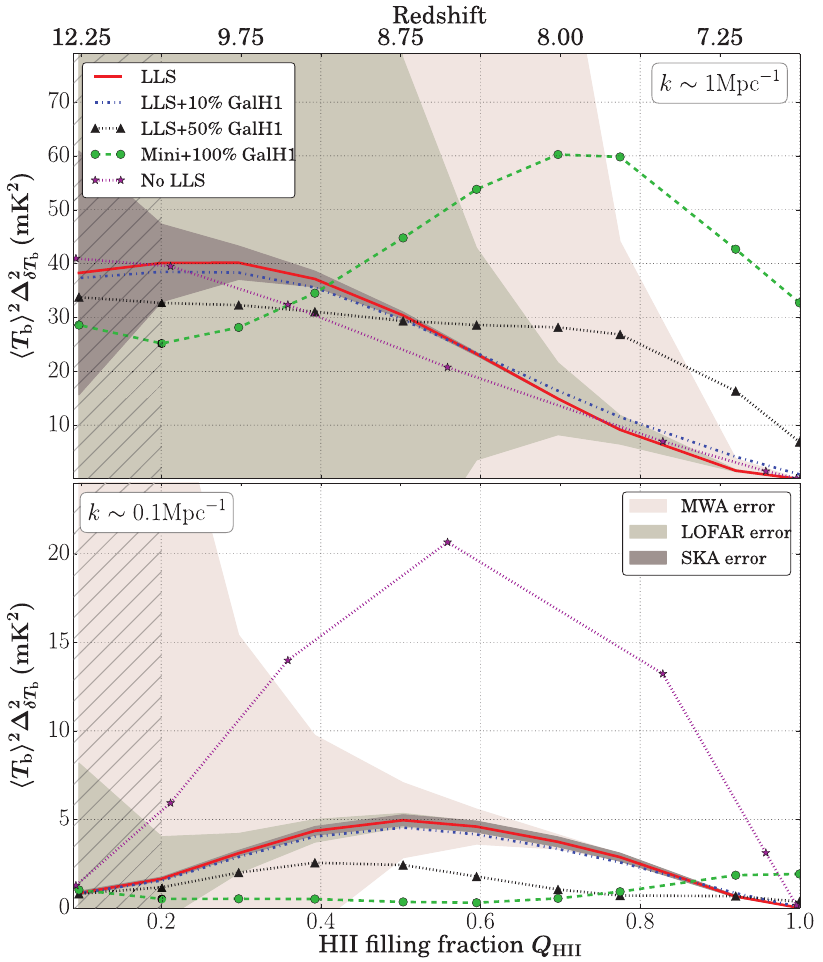} \\
  \caption{Dimensional power spectrum of the brightness temperature
as a function of the \hii volume-filling factor for $k=1\,$Mpc$^{-1}$ (top) and $k=0.1\,$Mpc$^{-1}$ (bottom). Within the hatched region, spin-temperature fluctuations may not be negligible.}
  \label{fig:PS_v_z}
\end{figure}

The evolution of the
dimensional power spectrum 
$\delta T^2_{\rm b}(\bmath{k},z)\,\Delta^2_{\delta T_{\rm b}}(k,z)$ 
as a function of the H\thinspace\textsc{ii} volume
filling factor is
plotted in Figure \ref{fig:PS_v_z} 
for $k=1$Mpc$^{-1}$ (top) and $k=0.1$Mpc$^{-1}$ (bottom).
Unlike higher-order moments, the power spectrum at small $k$
can cleanly pick-up large-scale structure.  As such 
its timing signatures are 
more robust.
For example, the mid-point turnover 
in the $\boldsymbol{k=0.1\textrm{Mpc}^{-1}}$ dimensional power spectrum
is robustly at
$Q_{ \mathrm{\textsc{hii}} }\sim 0.5$ for the 
more reasonable models, although it shifts to smaller
$Q_{ \mathrm{\textsc{hii}} }$  
with large levels of
galactic sinks 
and is completely wiped out in 
the extreme galactic \hi model. 
We see that whilst reasonable quantities of 
galactic sinks have little qualitative impact, large
quantities do impact on the
nature of the power spectrum's evolution with
its evolution no longer well described by 
a simple inverted parabola on large scales. 
This results from additional small-scale power
(evident in the plots of Figure \ref{fig:PS_v_k}
and the top plot of Figure \ref{fig:PS_v_z})
and from the fact that at late times there is 
still substantial residual \hi that drives the 
power.

\section{Concluding remarks}\label{sec:conc}

In this paper we have examined the impact
of sinks on the 21-cm statistics of reionization.
We consider two classes of sink, 
IGM sinks and galactic sinks,
which have two effects on the 21-cm signal:
\begin{packed_enum}
\item As absorbers of ionizing photons they impact on both the
timing of reionization and the morphology of cosmic \hii patches.
\item As reservoirs of neutral gas they contribute a 
non-zero 21-cm signal in otherwise ionized cosmic patches.
\end{packed_enum}
We use the simulations of \citet{Sobacchi2014} 
to investigate the impact of IGM sinks 
on the 21-cm 1-point statistics 
due to effects (i) and (ii).
In addition we study how the contribution
of galactic sinks to (ii) 
affects the 21-cm statistics. 
We also consider an extreme, illustrative model in which gas inside sterile minihaloes and atomically-cooled galaxies is fully neutral.

It is likely that the impact of sub-grid IGM sinks on reionization morphology (\citealt{Sobacchi2014}; effect i), will suppress the expected `rise and fall' reionization signal of the raw (unfiltered) 21-cm variance.
Instead, the large-scale power (or smoothed variance) shows a 
more pronounced evolution during reionization, 
since it is less affected by unresolved \hii structure.  
Thus first-generation instruments such as LOFAR should 
target large-scale modes for cleaner reionization probes.

Similarly, the rapid increase of the skewness
previously associated with the end of reionization
is also suppressed by inhomogeneous sinks.
This is because the variance
is non-zero even post-reionization (due to residual H\thinspace\textsc{i};
effect ii) and
so we instead see a turnover as the skewness transitions
from being dominated by the ionization field to the residual \hi field.

We find that more reasonable levels of galactic sinks,
will not change the qualitative evolution of the moments
during reionization. It is worth noting that the presence 
of even low levels ($<10$\%) of galactic sinks would result
in the absence of any late-time skewness signature in SKA maps. 
However, the late-time turnover in the skewness should be recoverable
by smoothing the maps to a lower resolution.

We find that large levels ($\gsim 50$\%) of galactic \hi 
have the potential to dramatically alter the qualitative
evolution of moments.
This provides an exciting opportunity for 
SKA (and even LOFAR) to constrain levels of galactic H\thinspace{\textsc i} in the high-redshift Universe.

\section*{Acknowledgements}
CW is supported by an STFC studentship. JRP acknowledges support under FP7-PEOPLE-2012-CIG grant \#321933-21ALPHA and STFC consolidated grant ST/K001051/1.

\bibliographystyle{mn2e}


\begin{thebibliography}{}
 \providecommand{\href}[2]{#2}
  \providecommand{\doi}[1]{\href{http://dx.doi.org/#1}{doi:#1}}
  \providecommand{\eprint}[1]{\href{http://arxiv.org/abs/#1}{arXiv:#1}}
\bibitem[\protect\citeauthoryear{Barkana \& Loeb}{Barkana \& Loeb}{2004}]{Barkana2004} Barkana R., Loeb A., 2004, ApJ, 609, 474

\bibitem[\protect\citeauthoryear{Baek, {Di Matteo}, Semelin, Combes \&
  Revaz}{Baek et~al.}{2009}]{Baek2009}
Baek S.,  {Di Matteo} P.,  Semelin B.,  Combes F.,    Revaz Y.,  2009, A\&A, 495, 389

\bibitem[\protect\citeauthoryear{Bennett et~al.,}{Bennett
  et~al.}{2013}]{Bennett2013}
Bennett C.~L.  et~al., 2013, ApJS, 208, 20

\bibitem[\protect\citeauthoryear{Bond, Cole, Efstathiou \& Kaiser}{Bond
  et~al.}{1991}]{Bond1991}
Bond J.~R.,  Cole S.,  Efstathiou G.,    Kaiser N.,  1991, ApJ, 379,
  440

\bibitem[\protect\citeauthoryear{Bolton 
\& Becker}{Bolton 
\& Becker}{2009}]{Bolton2009} Bolton J.~S., Becker G.~D., 2009, MNRAS, 398, L26

\bibitem[\protect\citeauthoryear{Choudhury, Ferrara \& Gallerani}{Choudhury
  et~al.}{2008}]{Choudhury2008}
Choudhury T.~R.,  Ferrara A.,    Gallerani S.,  2008, MNRAS Lett., 385, L58

\bibitem[\protect\citeauthoryear{Choudhury, Haehnelt \& Regan}{Choudhury
  et~al.}{2009}]{Choudhury2009}
Choudhury T.~R.,  Haehnelt M.~G.,    Regan J.,  2009, MNRAS, 394, 960

\bibitem[\protect\citeauthoryear{Ciardi, Scannapieco, Stoehr, Ferrara, Iliev \&
  Shapiro}{Ciardi et~al.}{2006}]{Ciardi2006}
Ciardi B.,  Scannapieco E.,  Stoehr F.,  Ferrara A.,  Iliev I.~T.,    Shapiro
  P.~R.,  2006, MNRAS, 366, 689

\bibitem[\protect\citeauthoryear{Crociani et 
al.}{Crociani et 
al.}{2011}]{Crociani2011a} Crociani D., Mesinger A., Moscardini L., 
Furlanetto S., 2011, MNRAS, 411, 289 

\bibitem[\protect\citeauthoryear{Fernandez \& Shull}{Fernandez \&
  Shull}{2011}]{Fernandez2011}
Fernandez E.~R.,  Shull J.~M.,  2011, ApJ, 731, 20

\bibitem[\protect\citeauthoryear{Furlanetto \& Oh}{Furlanetto \& Oh}
{2005}]{Furlanetto2005} Furlanetto S.~R., Oh S.~P., 2005, MNRAS, 363, 1031 


\bibitem[\protect\citeauthoryear{Furlanetto, Zaldarriaga \&
  Hernquist}{Furlanetto et~al.}{2004}]{Furlanetto2004}
Furlanetto S.~R.,  Zaldarriaga M.,    Hernquist L.,  2004, ApJ, 613,
  16

\bibitem[\protect\citeauthoryear{Ghara, Choudhury \& Datta}{Ghara
  et~al.}{2014}]{Ghara2014}
Ghara R.,  Choudhury T.,    Datta K.,  2014, preprint, \eprint{1406.4157}

\bibitem[\protect\citeauthoryear{Gnedin, Kravtsov \& Chen}{Gnedin
  et~al.}{2008}]{Gnedin2008}
Gnedin N.~Y.,  Kravtsov A.~V.,    Chen H.,  2008, ApJ, 672, 765

\bibitem[\protect\citeauthoryear{Haiman, Rees \& Loeb}{Haiman
  et~al.}{1997}]{Haiman1997}
Haiman Z.,  Rees M.~J.,    Loeb A.,  1997, ApJ, 476, 458

\bibitem[\protect\citeauthoryear{Haiman, Abel, 
\& Rees}{Haiman, Abel, 
\& Rees}{2000}]{Haiman2000a} Haiman Z., Abel T., Rees M.~J., 2000, ApJ, 534, 11 

\bibitem[\protect\citeauthoryear{Harker et~al.,}{Harker
  et~al.}{2009}]{Harker2009}
Harker G. J.~A.  et~al., 2009, MNRAS, 393, 1449

\bibitem[\protect\citeauthoryear{Iliev, Mellema, Ahn, Shapiro, Mao \&
  Pen}{Iliev et~al.}{2014}]{Iliev2013}
Iliev I.~T.,  Mellema G.,  Ahn K.,  Shapiro P.~R.,  Mao Y.,    Pen U.~L.,
  2014, MNRAS, 439, 725

\bibitem[\protect\citeauthoryear{Iliev, Shapiro \& Raga}{Iliev
  et~al.}{2005}]{Iliev2005}
Iliev I.~T.,  Shapiro P.~R.,    Raga A.~C.,  2005, MNRAS,
  361, 405

\bibitem[\protect\citeauthoryear{Jenkins, Frenk, White, Colberg, Cole, Evrard,
  Couchman \& Yoshida}{Jenkins et~al.}{2001}]{Jenkins2001}
Jenkins A.,  Frenk C.~S.,  White S. D.~M.,  Colberg J.~M.,  Cole S.,  Evrard
  A.~E.,  Couchman H. M.~P.,    Yoshida N.,  2001, MNRAS,
  321, 372

\bibitem[\protect\citeauthoryear{Lacey \& Cole}{Lacey \&
  Cole}{1993}]{LaceyCedric1993}
Lacey C.,  Cole S.,  1993, MNRAS, 262, 627

\bibitem[\protect\citeauthoryear{Lidz et al.}{Lidz et al.}{2008}]{Lidz2008} 
Lidz A., Zahn O., McQuinn M., Zaldarriaga M., Hernquist L., 2008, ApJ, 680, 
962 

\bibitem[\protect\citeauthoryear{McQuinn, Zahn, Zaldarriaga, Hernquist \&
  Furlanetto}{McQuinn et~al.}{2006}]{McQuinn2006}
McQuinn M.,  Zahn O.,  Zaldarriaga M.,  Hernquist L.,    Furlanetto S.~R.,
  2006, ApJ 653, 815

\bibitem[\protect\citeauthoryear{McQuinn, Oh, 
\& Faucher-Gigu{\`e}re}{McQuinn, Oh, 
\& Faucher-Gigu{\`e}re}{2011}]{McQuinn2011} McQuinn M., Oh S.~P., Faucher-Gigu{\`e}re C.-A., 2011, ApJ, 743, 82 

\bibitem[\protect\citeauthoryear{Mellema et~al.,}{Mellema
  et~al.}{2013}]{Mellema2013}
Mellema G.  et~al., 2013, Exp. Astron., 36, 235

\bibitem[\protect\citeauthoryear{Mesinger, Bryan, 
\& Haiman}{Mesinger, Bryan, 
\& Haiman}{2006}]{Mesinger2006} Mesinger A., Bryan G.~L., Haiman Z., 2006, ApJ, 648, 835 

\bibitem[\protect\citeauthoryear{Mesinger \& Furlanetto}{Mesinger 
\& Furlanetto}{2007}]{Mesinger2007} Mesinger A., Furlanetto S., 2007, ApJ, 669, 663 

\bibitem[\protect\citeauthoryear{Mesinger \& Dijkstra}{Mesinger \&
  Dijkstra}{2008}]{Mesinger2008}
Mesinger A.,  Dijkstra M.,  2008, MNRAS, 390, 1071

\bibitem[\protect\citeauthoryear{Mesinger, Furlanetto \& Cen}{Mesinger
  et~al.}{2011}]{Signal2010}
Mesinger A.,  Furlanetto S.~R.,    Cen R.,  2011, MNRAS,
  411, 955

\bibitem[\protect\citeauthoryear{Mesinger, Ferrara \& Spiegel}{Mesinger
  et~al.}{2013}]{Mesinger2013a}
Mesinger A.,  Ferrara A.,    Spiegel D.~S.,  2013, MNRAS,
  431, 621

\bibitem[\protect\citeauthoryear{Miralda-Escude}{Miralda-Escude}{2003}]{Mirald%
aEscude2003}
Miralda-Escude J.,  2003, ApJ, 597, 66

\bibitem[\protect\citeauthoryear{Miralda-Escude, Haehnelt \&
  Rees}{Miralda-Escude et~al.}{2000}]{Miralda-Escude2000}
Miralda-Escude J.,  Haehnelt M.,    Rees M.~J.,  2000, ApJ, 530, 1

\bibitem[\protect\citeauthoryear{Mu{\~n}oz et 
al.}{Mu{\~n}oz et al.}{2014}]{Munoz2014} 
Mu{\~n}oz J.~A., Oh S.~P., Davies F.~B., 
Furlanetto S.~R., 2014, preprint, \eprint{1410.2249} 

\bibitem[\protect\citeauthoryear{Noterdaeme et~al.,}{Noterdaeme
  et~al.}{2012}]{Noterdaeme2012}
Noterdaeme P.  et~al., 2012, A\&A, 547, L1

\bibitem[\protect\citeauthoryear{O'Meara, Prochaska, Worseck, Chen \&
  Madau}{OâMeara et~al.}{2013}]{OMeara2013a}
O'Meara J.~M.,  Prochaska J.~X.,  Worseck G.,  Chen H.~W.,   Madau P.,
  2013, ApJ, 765, 137

\bibitem[\protect\citeauthoryear{Pober et~al.,}{Pober et~al.}{2014}]{Pober2013}
Pober J.~C.  et~al., 2014, ApJ, 782, 66

\bibitem[\protect\citeauthoryear{Pritchard \& Furlanetto}{Pritchard \&
  Furlanetto}{2007}]{Pritchard2007}
Pritchard J.~R.,  Furlanetto S.~R.,  2007, MNRAS, 376,
  1680

\bibitem[\protect\citeauthoryear{Prochaska, HerbertâFort \& Wolfe}{Prochaska
  et~al.}{2005}]{Prochaska2005}
Prochaska J.~X.,  HerbertâFort S.,    Wolfe A.~M.,  2005, ApJ, 635,
  123

\bibitem[\protect\citeauthoryear{Rahmati, Pawlik, Raicevic \& Schaye}{Rahmati
  et~al.}{2013}]{Rahmati2013}
Rahmati A.,  Pawlik A.~H.,  Raicevic M.,    Schaye J.,  2013, MNRAS, 430, 2427

\bibitem[\protect\citeauthoryear{Ricotti, Gnedin, 
\& Shull}{Ricotti, Gnedin, 
\& Shull}{2001}]{Ricotti2001} Ricotti M., Gnedin N.~Y., Shull J.~M., 2001, ApJ, 560, 580 

\bibitem[\protect\citeauthoryear{Santos, Ferramacho, Silva, Amblard, Cooray \&
  Santos}{Santos et~al.}{2010}]{Santos2010}
Santos M.~G.,  Ferramacho L.,  Silva M.~B.,  Amblard A.,  Cooray A.,    Santos
  G.,  2010, MNRAS, 406, 2421

\bibitem[\protect\citeauthoryear{Schaye}{Schaye}{2001}]{Schaye2001}
Schaye J.,  2001, ApJ, 559, 507

\bibitem[\protect\citeauthoryear{Shapiro, Iliev \& Raga}{Shapiro
  et~al.}{2004}]{Shapiro2004}
Shapiro P.~R.,  Iliev I.~T.,    Raga A.~C.,  2004, MNRAS,
  348, 753

\bibitem[\protect\citeauthoryear{Sheth \& Tormen}{Sheth \&
  Tormen}{1999}]{Sheth1999}
Sheth R.~K.,  Tormen G.,  1999, MNRAS, 308, 119

\bibitem[\protect\citeauthoryear{Sobacchi 
\& Mesinger}{Sobacchi 
\& Mesinger}{2013}]{Sobacchi2013} Sobacchi E., Mesinger A., 2013, MNRAS, 432, L51 

\bibitem[\protect\citeauthoryear{Sobacchi \& Mesinger}{Sobacchi \&
  Mesinger}{2014}]{Sobacchi2014}
Sobacchi E.,  Mesinger A.,  2014, MNRAS, 440, 1662

\bibitem[\protect\citeauthoryear{Songaila \& Cowie}{Songaila \&
  Cowie}{2010}]{Songaila2010a}
Songaila A.,  Cowie L.~L.,  2010, ApJ, 721, 1448

\bibitem[\protect\citeauthoryear{Storrie-Lombardi, McMahon, Irwin \&
  Hazard}{Storrie-Lombardi et~al.}{1994}]{Storrie-Lombardi1994}
Storrie-Lombardi L.~J.,  McMahon R.~G.,  Irwin M.~J.,    Hazard C.,  1994,
  ApJ, 427, L13

\bibitem[\protect\citeauthoryear{Thomas \& Zaroubi}{Thomas \&
  Zaroubi}{2011}]{Thomas2011}
Thomas R.~M.,  Zaroubi S.,  2011, MNRAS, 410, 1377

\bibitem[\protect\citeauthoryear{Thoul \& Weinberg}{Thoul \&
  Weinberg}{1996}]{Thoul1996}
Thoul A.~A.,  Weinberg D.~H.,  1996, ApJ, 465, 608

\bibitem[\protect\citeauthoryear{Tingay et~al.,}{Tingay
  et~al.}{2013}]{Tingay2013}
Tingay S.~J.  et~al., 2013, Publ. Astron. Soc. Aust., 30, 1

\bibitem[\protect\citeauthoryear{Trac \& Cen}{Trac 
\& Cen}{2007}]{Trac2007} Trac H., Cen R., 2007, ApJ, 671, 1 

\bibitem[\protect\citeauthoryear{Trac 
\& Gnedin}{Trac 
\& Gnedin}{2011}]{Trac2011} Trac H.~Y., Gnedin N.~Y., 2011, ASL, 4, 228 

\bibitem[\protect\citeauthoryear{Watkinson \& Pritchard}{Watkinson \&
  Pritchard}{2014}]{Watkinson2014}
Watkinson C.~A.,  Pritchard J.~R.,  2014, MNRAS, 443, 3090

\bibitem[\protect\citeauthoryear{Wolfe, Gawiser, 
\& Prochaska}{Wolfe, Gawiser, 
\& Prochaska}{2005}]{Wolfe2005} Wolfe A.~M., Gawiser E., Prochaska J.~X., 2005, ARA\&A, 43, 861 

\bibitem[\protect\citeauthoryear{Worseck \& Prochaska}{Worseck \&
  Prochaska}{2011}]{Worseck2011}
Worseck G.,  Prochaska J.~X.,  2011, ApJ, 728, 23

\bibitem[\protect\citeauthoryear{Wyithe, Warszawski, Geil \& Oh}{Wyithe
  et~al.}{2009}]{Wyithe2009}
Wyithe J. S.~B.,  Warszawski L.,  Geil P.~M.,    Oh S.~P.,  2009, MNRAS, 395, 311

\bibitem[\protect\citeauthoryear{Yajima, Choi \& Nagamine}{Yajima
  et~al.}{2011}]{Yajima2011}
Yajima H.,  Choi J.-H.,    Nagamine K.,  2011, MNRAS, 412,
  411

\bibitem[\protect\citeauthoryear{{Planck Collaboration} et~al.,}{{Planck
  Collaboration} et~al.}{2014}]{PlanckCollaboration2013b}
{Planck Collaboration} et~al., 2014, A\&A, 566, \eprint{1303.5076}

\bibitem[\protect\citeauthoryear{Zahn et~al.}{Zahn et~al.}{2007}]{Zahn2007} 
Zahn, O., Lidz, A., McQuinn, M., et al. 2007, ApJ, 654, 12 

\bibitem[\protect\citeauthoryear{Zahn et al.}{Zahn et al.}{2011}]{Zahn2011} 
Zahn O., Mesinger A., McQuinn M., Trac H., Cen R., Hernquist L.~E., 2011, 
MNRAS, 414, 727 

\bibitem[\protect\citeauthoryear{Zel'dovich}{Zel'dovich}{1970}]{Zel'dovichYa.B.1970} Zel'dovich Y.~B., 1970, A\&A, 5, 84 
\end{thebibliography}

\bsp

\end{document}